\documentclass[useAMS,usenatbib]{mn2e}

\bibpunct{(}{)}{;}{a}{}{,}
\usepackage{graphicx}

\def\vsini{$V\!\sin i$}
\def\teff{T$_ {\rm{eff}}$}
\def\logg{log~{\it g}}

\def\kps{km~s$^{\rm{-1}}$}
\DeclareRobustCommand{\ion}[2]{%
\relax\ifmmode \ifx\testbx\f@series
{\mathbf{#1\,\mathsc{#2}}}\else {\mathrm{#1\,\mathsc{#2}}}\fi
\else\textup{#1\,{\mdseries\textsc{#2}}}%
\fi}

\title[Spectral disentangling of the triple system DG Leo]{Spectral disentangling of the triple system DG Leo: orbits and chemical composition\thanks{This work is based on observations made at the Haute-Provence Observatory (France).}}
\author[Y. Fr\'emat, P. Lampens \& H. Hensberge]{Y. Fr\'emat, P. Lampens, H. Hensberge\\
Royal Observatory of Belgium, 3 ringlaan, B-1180 Brussel
(Belgium)}
\begin{document}

\date{Accepted: October 7, 2004}

\pagerange{\pageref{firstpage}--\pageref{lastpage}} \pubyear{2004}

\maketitle

\label{firstpage}

\begin{abstract}
DG Leo is a spectroscopic triple system composed by 3 stars of
late-A spectral type, one of which was suggested to be a $\delta$
Scuti star. Seven nights of observations at high spectral and high
time resolution at the Observatoire de Haute-Provence with the
ELODIE spectrograph were used to obtain the component spectra by
applying a Fourier transform spectral disentangling technique.
Comparing these with synthetic spectra, the stellar fundamental
parameters (effective temperature, surface gravity, projected
rotation velocity and chemical composition) are derived. The
inner binary consists of two Am components, at least one of which
is not yet rotating synchronously at the orbital period though the
orbit is a circular one. The distant third component is confirmed
to be a $\delta$ Scuti star with normal chemical composition.
\end{abstract}


\section{Introduction}

   Different reasons can be invoked to illustrate the fact that main-sequence A-type stars occupy a
   very interesting region of the H-R diagram:
\begin{itemize}
   \item[--] the transition from radiative to convective energy transport occurs at effective temperatures
   between 8500 and 6000 K (from A5 to F5), a range which encompasses the late A-type stars;
   \item[--] the classical Cepheid instability strip, when extended downward, intersects the ZAMS \citep[cf. Fig.~8 in][]{2001A&A...366..178R} at
   effective temperatures of 8800 K (blue edge) and 7500 K (red edge)
   (spectral types from A3 to F0);
   \item[--] the presence of magnetism is clearly demonstrated from the mid-B to early-F spectral types;
\item[--] metal-line abundance anomalies are also frequently detected among the (non-magnetic) late A-type stars.
\end{itemize}
   The variety of (atmospheric) phenomena which may affect the stars located in that part of the H-R
  diagram in one or the other way is extremely rich: these include several possible pulsation driving
  mechanisms (acting in the $\delta$ Scuti and SX Phe, $\gamma$ Dor or roAp variable stars) and different
  processes of magnetism, diffusion, rotation and convection which are thought to boost or inhibit the
  presence of chemical peculiarities in the stellar atmospheres (Ap, Am, $\rho$ Pup and $\lambda$
  Boo stars). Among the pulsators there exist: the $\delta$ Scuti stars which are main-sequence or giant
  low-amplitude variable stars pulsating in radial and non-radial acoustic modes (p modes) with typical
  periods of 0.02--0.25 days; the SX Phe stars which are high-amplitude $\delta$ Scuti stars
  of the old disk population \citep{2000dsrs.conf....3B}; the $\gamma$ Dor stars which are cooler and pulsate in
  non-radial gravity modes (g modes) with typical periods of 1--2 days \citep{2000dsrs.conf....3B}; rapidly
  oscillating cooler magnetic Ap (roAp) stars that exhibit very high-overtone non-radial acoustic pulsations
  aligned with the magnetic axis, inclined relatively to the
  rotation axis (`oblique pulsator' model) and with typical variability periods of 5--16 min \citep{2000dsrs.conf..287K}. Among the chemically
  peculiar (CP) stars of interest, we have the magnetic Ap stars that are generally variable owing to
  surface inhomogeneities coupled to rotation (model known as the `oblique rotator'); the non-magnetic
  metallic-lined stars such as the classical Am stars which have K-line and metal-line spectral types
  differing by at least 5 spectral subclasses, and the evolved Am stars of luminosity class IV and III
  called $\rho$ Puppis \citep[also $\delta$ Delphini stars,][]{2000dsrs.conf..287K}; and the $\lambda$ Bootis stars which
  show weak metal lines owing to underabundances of the Fe-peak elements \citep{2000dsrs.conf..287K}. Some of
  these phenomena may or may not be mutually exclusive. For example, enhanced metallicity and pulsation
  can theoretically occur together in the cooler and/or evolved part of the instability strip while there
  is mutual exclusion for the hotter A-stars of the main--sequence \citep{2000A&A...360..603T}. The role of
  multiplicity is an additional aspect that has to be considered, as the majority of the Am stars are
  binaries with orbital periods between 1 and 10 days \citep{1996A&A...313..523B,1997A&A...326..655B}. Most of these
  binaries are tidally locked systems with synchronized orbital and rotational periods. Their projected
  equatorial velocities are generally below 100 \kps~\citep{2000dsrs.conf..287K}. Because of this tidal braking mechanism,
  the process of diffusion can act more efficiently. In the case of the (magnetic) Ap stars, the braking
  mechanism is magnetic. Their projected equatorial velocities are usually well below 100 \kps~\citep{2000dsrs.conf..287K}.
  Thus rotation is relevant for explaining the presence of chemical peculiarities. An unknown multiplicity status is further
  expected to be the cause of some continuum veiling in several $\lambda$
  Bootis stars leading to an underestimation of the abundance of
  the Fe-peak elements as was recently suggested in \citet{2003A&A...412..447G} and \citet{2003A&A...398..697F}. It also appears that about 50\% of
  these exhibit pulsations of $\delta$ Scuti type \citep{2000dsrs.conf..410W}. Thus complex interactions
  between all of these processes exist which are not easily disentangled.

For all these reasons the analysis of the chemical composition of
multiple systems having at least one pulsating component is an
ideal tool to explore in an empirical way the interactions that
may or may not exist between pulsation, diffusion, rotation and
multiplicity. In the present paper, we are therefore dealing with
the chemical analysis of DG Leo (HD~85040, HR~3889, HIP~48218,
Kui~44), a known multiple system with a pulsating component that
we describe in Sect.~\ref{sec:dgleo}. Details about the
spectroscopic observations we performed are given in
Sect.\ref{sec:observations} while the Fourier-transform technique
that was adopted to disentangle the component spectra from the
combined ones is described in Sect.~\ref{sec:korel}. Orbital
parameters of both spectroscopic and visual orbits are discussed
in Sect.~\ref{sec:aorb}. Fundamental parameters and chemical
composition of all three components are derived in
Sect.~\ref{sec:fundpar} and Sect.~\ref{sec:abundance analysis}
respectively.


\begin{figure}
\center
\begin{minipage}{8cm}
\center
\includegraphics[width=7cm,angle=0,clip=]{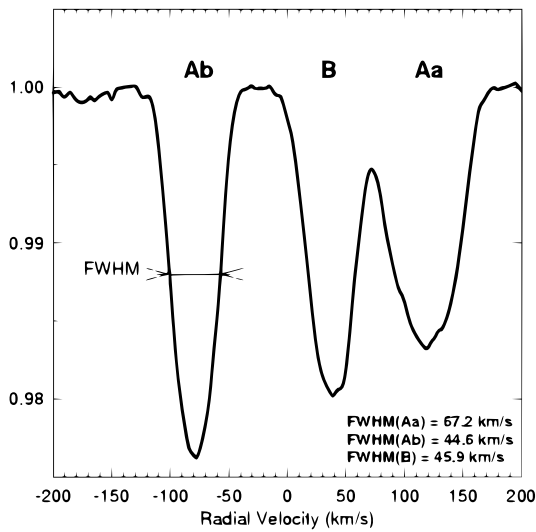}
\caption{Two hours-averaged cross correlation function observed
from JD2452651.6036 to JD2452651.6954 (i.e.: orbital
phase~$\sim$~0.98), when all three components are resolved. Each
correlation peak belongs to one component of DG Leo (Aa, Ab or
B).} \label{fig:crossc}
\end{minipage}
\end{figure}

\section{DG Leo}

\label{sec:dgleo}

In this work we present a detailed spectroscopic study of the
hierarchical triple system DG~Leo, which consists of a close
binary (components Aa and Ab) and one distant companion
(component B) and whose components were classified as late A-type
stars. The orbital period of the Aa,b system is 4.15 days
\citep{1967ApJ...149...55D, 1977PASP...89..216F} while the
orbital period of the visual pair AB was estimated to be roughly
200 years \citep{1977PASP...89..216F}. Both close components have
spectral type A8~IV \citep{1982bsc..book.....H} while the
composite spectrum was classified as being of type F0~IIIn
\citep{1976PASP...88...95C} and of type A7~III with enhanced Sr
\citep{1979PASP...91...83C}. All three components of the system
are located 
in the $\delta$ Scuti instability strip and are therefore
potential candidates for pulsations. However, the claims for
short-period oscillations seem mainly to concern one component:
component B was classified both as an ultra-short-period Cepheid
\citep{1979ApJS...41..413E} and as a $\delta$ Scuti star
\citep{1974AJ.....79.1082E}. 
As a matter of fact, multiple short-period oscillations of
$\delta$ Scuti type with periodicities of about 2 hrs
\citep{phot1} have been detected in the combined light curve.

\section{Observations and data reduction}

\label{sec:observations}

The triple system is spectroscopically unresolved, but the
Doppler information available in the composite spectrum can be
used to recover the contributions of the individual components.
Therefore, extensive spectroscopy was obtained at the
Haute-Provence Observatory (OHP) with the ELODIE spectrograph
\citep{1996A&AS..119..373B} on the 1.93-m telescope. Observations
were collected during 7 nights in 2003 (January 3 -- 7 \& January
11 -- 15) and cover about 50 percent of the close binary's
(component A) orbital phase. The time exposure was fixed at 360
seconds in order to resolve the presumed p-mode pulsations, yet
with a good signal-to-noise ratio and a high resolution (about
50000).

The data were automatically reduced order by order with the
INTERTACOS pipeline \citep{1996A&AS..119..373B} at the end of
each night. This reduction procedure takes care of the order
extraction, of the offset and flat-field corrections and of the
wavelength calibration using a thorium reference spectrum. The
resulting wavelength scale is corrected for earth motion
afterwards by means of the IRAF software package. In this way, 245
spectra ranging from 3900 to 6800~\AA~were obtained with a
signal-to-noise ratio that generally varies between 100 to 180 at
5500 \AA.

\begin{table*}
\begin{minipage}{12cm}
\center \caption[]{Sample of the radial velocities measured on the
cross-correlation functions (the complete electronic version is
available at the CDS). Column 1 gives the Julian date of the
observations, columns 2, 4 and 6 list the radial velocities of
components B, Ab and Aa respectively. Errors on radial-velocity
measurements (column 3, 5 and 7) are the r.m.s. computed by
fitting the cross-correlation peak with a Gaussian profile over
three different intervals.} \label{tab:vrccf}
\begin{tabular}{lrrrrrr}
\hline
JD$-$2400000  & V$\rm{B}$      &  $\sigma_{\rm{B}}$    &
V$\rm{Ab}$       & $\sigma_{\rm{Ab}}$     & V$\rm{Aa}$       &
$\sigma_{\rm{Aa}}$   \\\vspace{0.5mm}
            & [km s$^{\rm{-1}}$] & [km s$^{\rm{-1}}$] & [km s$^{\rm{-1}}$] & [km s$^{\rm{-1}}$] & [km s$^{\rm{-1}}$] & [km s$^{\rm{-1}}$] \\
\hline
...\\
52651.4773 &   33.48 & 0.52 & $-$74.64 & 0.26 & 115.71 & 0.56\\
52651.4830 &   33.16 & 0.15 & $-$76.31 & 0.10 & 112.27 & 0.22\\
52651.4886 &   35.12 & 0.19 & $-$74.87 & 0.05 & 112.14 & 0.32\\
52651.4943 &   35.47 & 0.18 & $-$74.60 & 0.67 & 110.59 & 0.37\\
52651.5000 &   38.14 & 0.35 & $-$74.95 & 0.09 & 118.18 & 0.28\\
52651.5057 &   38.61 & 0.33 & $-$76.75 & 0.06 & 117.81 & 0.56\\
52651.5114 &   39.27 & 0.22 & $-$75.58 & 0.09 & 115.21 & 3.10\\
52651.5172 &   39.53 & 0.27 & $-$75.56 & 0.12 & 114.48 & 0.34\\
52651.5229 &   39.71 & 0.53 & $-$75.46 & 0.10 & 114.09 & 0.20\\
52651.5292 &   40.06 & 0.68 & $-$75.66 & 0.15 & 110.95 & 0.74\\
...\\
\hline
\end{tabular}
\end{minipage}
\end{table*}
\begin{table*}
\begin{minipage}{12cm}
\center \caption[]{Sample of radial velocities derived with {\sc
korel} (the complete electronic version is available at the CDS).
Column 1 gives the Julian date of the observations, columns 2, 4
and 6 list the radial velocities of components B, Ab and Aa
respectively. The deviations for each measurement relatively to
the orbital solution, or (O$-$C)s, are given in columns 3, 5 and
7.} \label{tab:vrkorel}
\begin{tabular}{lrrrrrr}
\hline JD$-$2400000  & V$\rm{B}$      &  (O$-$C)    &
V$\rm{Ab}$       & (O$-$C)     & V$\rm{Aa}$       &  (O$-$C)
\\\vspace{0.5mm}
            & [km s$^{\rm{-1}}$] & [km s$^{\rm{-1}}$] & [km s$^{\rm{-1}}$] & [km s$^{\rm{-1}}$] & [km s$^{\rm{-1}}$] & [km s$^{\rm{-1}}$]
            \\\hline
...\\
52651.4773 &   37.60 & $-$0.87 &  $-$75.01 & $-$0.52 &  112.56 & $-$2.30 \\
52651.4830 &   36.71 & $-$1.76 &  $-$76.01 & $-$1.23 &  113.49 & $-$1.66 \\
52651.4886 &   36.90 & $-$1.57 &  $-$75.90 & $-$0.84 &  114.02 & $-$1.41 \\
52651.4943 &   37.34 & $-$1.14 &  $-$76.75 & $-$1.40 &  115.45 & $-$0.25 \\
52651.5000 &   37.62 & $-$0.85 &  $-$76.13 & $-$0.51 &  114.74 & $-$1.24 \\
52651.5057 &   38.28 & $-$0.19 &  $-$76.72 & $-$0.83 &  115.49 & $-$0.75 \\
52651.5114 &   38.47 &  0.00 &  $-$77.49 & $-$1.35 &  113.88 & $-$2.62 \\
52651.5172 &   38.83 &  0.36 &  $-$76.73 & $-$0.34 &  115.73 & $-$1.02 \\
52651.5229 &   39.14 &  0.67 &  $-$77.07 & $-$0.43 &  115.53 & $-$1.47 \\
52651.5292 &   38.47 &  0.00 &  $-$77.20 & $-$0.30 &  114.75 & $-$2.51 \\
...\\ \hline
\end{tabular}
\end{minipage}
\end{table*}

After each exposure, INTERTACOS provided us also with a
correlation function (see Fig.~\ref{fig:crossc}) computed using a
template corresponding to a F0 V type star and that accounts for
about 2000 spectral lines. This function allowed us to measure
accurately the heliocentric radial velocity of DG Leo's
components, but only at those orbital phases where the component
structure in the cross-correlation function is sufficiently
resolved. A sample of the radial velocities measured in this way
and at these specific phases is listed in Table \ref{tab:vrccf}.
An electronic version of this table is available at the Centre de
Donn\'ees Stellaires de Strasbourg (CDS).

\section{Spectral disentangling}

\label{sec:korel}

{The composite spectra of DG Leo are complex, and a careful
quantitative analysis is needed. To extract the individual
contributions of the three components, we adopted the spectral
disentangling technique. This technique determines the
contributions of the components to the composite spectra and the
orbital parameters in a self-consistent way. The feasibility of
the disentangling method in practice was first proved by
\citet{1994A&A...281..286S}. They reconstructed the spectrum of
the primary component of V453 Cygni from composite spectra at
different orbital phases out of eclipse and showed that it
matched perfectly the one observed during total eclipse in the
secondary minimum. Since then, different implementations of the
method have been successfully applied to binaries with components
ranging from O-type \citep{1994A&A...282...93S} to F-type
components \citep{2002AJ....123..988G}. Advantages and
disadvantages of several implementations have been discussed by
\citet{Ilijic}.

We used the technique as developed by \citet[][and references
therein]{1995A&AS..114..393H} and applied in the {\sc korel}
computer code (Release 21.3.99). Although the method no longer
requires the intermediate step of deriving radial velocities,
{\sc korel} cross-correlates the input spectra with the
disentangled component spectra afterwards to provide relative
radial velocities for comparison with historical data. Note that
the systemic velocity must be determined independently, as the
concept of spectral lines does not enter in the purely
mathematical disentangling procedure. A basic assumption on which
the disentangling is carried out is that the shape of the line
profiles of the binary components should not vary with time.
However, the technique has been found to be applicable in the
presence of complex variations of line shapes under certain
conditions \citep{2004A&A...422.1013H,decat}. Helpful conditions
are: a very different time-scale for line variability and orbital
revolution; many spectra and line variability not phase-locked
with the orbital motion; amplitude of line variability small with
respect to the weakest contributing component. Under these
conditions the variations simply act as additional high-frequency
noise. We are in a relatively favourable case in all aspects and
especially because the three components each contribute roughly
one third to the composite spectrum while line variability will
turn out to be moderate and limited to only one component. All
245 spectra obtained at {\sc ohp} were used in the disentangling
procedure.}

\subsection{Component spectra and luminosity ratios}

When the components' relative fluxes are known or when they vary
substantially with time \citep[e.g. when eclipses are
observed,][]{2000A&A...358..553H}, the method does allow to
account straightforwardly for the veiling owing to multiplicity
which artificially weakens all the spectral lines.
\citet{ilijic04} show that in the case of time-independent
component fluxes, the spectral disentangling can be performed
assuming equal fluxes and the resulting {\it spectra} can be
re-normalised afterwards. Therefore, we first assumed that each
component contributes equally to the total luminosity. Then, to
correct the resulting component spectra, we estimated the
relative luminosity at $\lambda$3930~\AA~with the assumption that
the Ca~{\sc ii}~K line depth is saturated (Sect.~\ref{sec:calr}).
We finally applied this ratio to the whole investigated
wavelength range in view of the similar flux distribution of the
3 stars (Table~\ref{tab:fund}). As a result, the luminosity
ratios and the estimated individual V magnitudes we derived are
given in Table \ref{tab:lr}. The implied difference between the
magnitudes of components A and B ($\Delta V = 0.59 \pm 0.09$) is
consistent with the differential magnitude measured by HIPPARCOS
\citep[$\Delta Hp=0.69 \pm 0.08$;][]{1997yCat.1239....0E}.
Examples of the resulting components' spectra can be found in
Fig.~\ref{fig:fit}. The complete disentangled spectra are
available at the CDS as ps-format files.

\begin{figure}
\center
\begin{minipage}{8cm}
\includegraphics[width=8cm,angle=0,clip=]{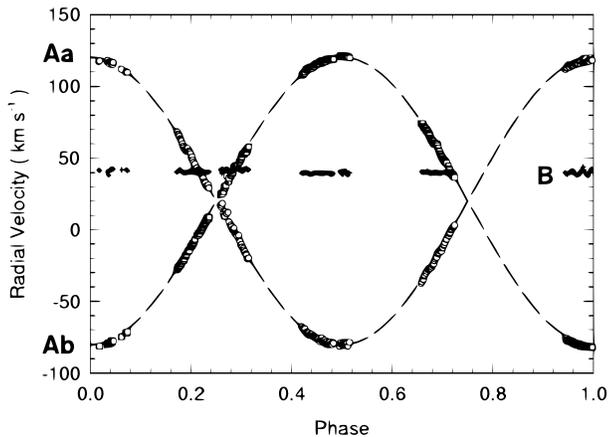}
\caption{Radial velocities obtained with {\sc korel} after the
disentangling procedure, and reported in the heliocentric
framework.} \label{fig:vrkor}
\end{minipage}
\end{figure}

\subsection{Radial Velocities}

As noted earlier, {\sc korel} cross-correlates the input
composite spectra with the output disentangled component spectra
afterwards in order to provide radial velocities with respect to
the centre of mass of the system. To estimate the systemic
velocity of the DG Leo system, the radial velocities deduced with
{\sc korel} were compared to those measured on a subset of
spectra with INTERTACOS ($\gamma$~=~26.54~$\pm$~0.5~\kps). A
sample of these {\sc korel}-based heliocentric radial velocities
is given in Table~\ref{tab:vrkorel}. The complete data set is
electronically available at the CDS and plotted in
Fig.~\ref{fig:vrkor} against the orbital phase of the close
binary (Aab system).

A detailed comparison of both types of radial-velocity
measurements (Tables~\ref{tab:vrccf} and \ref{tab:vrkorel}) in
the common subset shows no significant offset (at the
1~\kps~level) for any component, but a different scatter for the
three stars: the largest scatter (differences up to 10~\kps) is
observed for the pulsating (Sect.~8.6) B component, while the
smallest scatter (differences up to 3~\kps) applies to the Ab
component. The wider lines of Aa produce differences up to
6~\kps. The more realistic correlation masks from the {\sc korel}
procedure undoubtedly produce better radial-velocity estimates
than the Gaussian component fitting of the INTERTACOS CCFs (in
the sense of lower residuals relative to the orbital solution),
especially for the partly overlapping Aa and B components.
Therefore, the radial velocities used in the further analysis are
those derived for all 245 spectra with the component spectra as
correlation masks.

\begin{table}
\center
\begin{minipage}{6cm}
\center \caption{Luminosity ratios (r$_{\rm{L}}$) and component
magnitudes (V$_{\rm{i}}$) derived from the depth of the
\ion{Ca}{ii} K line. Note that errors are not uncorrelated, since
the combined magnitude is known accurately and the sum of the
fractional luminosities is exactly unity.} \label{tab:lr}
\begin{tabular}{lrr}
\hline Component                 & r$_{\rm{L}}$ & V$_{\rm{i}}$
                 \\\vspace{0.5mm}
                 &     (\%)      &    (mag)             \\
                 \hline
DG Leo Aa        &   32 $\pm$ 2  & 7.31 $\pm$ 0.15      \\
DG Leo Ab        &   31 $\pm$ 2  & 7.35 $\pm$ 0.15      \\
DG Leo B         &   37 $\pm$ 2  & 7.16 $\pm$ 0.14      \\
\hline
\end{tabular}

\end{minipage}
\end{table}


\section{{Analysis of the orbits}}

\label{sec:aorb}

\subsection{Spectroscopic orbit of the close binary Aab}

\label{subsec:OPSB}

The orbital parameters were derived applying {\sc korel} to all
245 spectra in 5 spectral regions. The regions were selected for
showing the largest flux gradients. These wavelength domains were
basically those having simultaneously the greatest line density
and the fewest blends. During the fitting procedure, 5 parameters
were finally used to describe the orbit of the close binary
system (A) and were defined as free parameters, while the
longitude of periastron was kept fixed (i.e.: $\omega=0^o$). The
final solution is:

\begin{center}
\begin{tabular}{rcl}
P &=& 4.146751~$\pm$~0.000005~\rm{days} \\
T$_{\rm{0}}$ &=& 2452639.259~$\pm$~0.001 \rm{JD} \\
e &=& 0.0000~$\pm$~0.0004\\
K$_{\rm{Aa}}$ &=& 100.72~$\pm$~0.18~km~s$^{\rm{-1}}$\\
q &=& 1.000~$\pm$~0.001
\end{tabular}
\end{center}
\noindent where P is the orbital period, T$_{\rm{0}}$ is the time
of nodal passage, e is the eccentricity, K$_{\rm{Aa}}$ is the
amplitude in radial velocity of component Aa and q is the mass
ratio. {Notice that, as the orbit is circular, the time of nodal
passage was adopted. Since {\sc korel} does not give error
estimates, uncertainties were estimated from the covariance
matrix using the radial-velocities as input. This is a
conservative way of estimating uncertainties, since it involves
the derived radial velocities, that are vulnerable to
cross-correlation bias in a composite spectrum. Actually,
plotting the radial-velocity residuals (listed in
Table~\ref{tab:vrkorel}) against phase indicates correlated
errors at the 1~\kps~level, with strings of characteristically 6
consecutive (O$-$C) having the same sign. This is taken into
account explicitly in the uncertainty on the velocity amplitude
by the method proposed by \citet{1998BaltA...7...43S}.
Furthermore, the accuracy of the orbital period was obtained by
taking into account the uniquely determined number of cycles
between our observations and those of \citet{1977PASP...89..216F}.
It is further worth noting that the period of the visual binary
(AB) is so long compared to the time coverage of our observations
that its orbit and motion can be neglected.

\subsection{Astrometric--spectroscopic orbit of the wide binary AB}

Although micrometric observations of DG~Leo exist since 1935
\citep{1935PASP...47..230K}, no orbit is known yet.
\citet{1977PASP...89..216F} roughly estimated an orbital period of
180 $\pm$ 50 yrs based on {calibrations of spectral type versus
mass and absolute magnitude}. Notice that in their study a
parallax of 0.007\arcsec was adopted which is in full concordance
with the Hipparcos parallax (6.34 $\pm$ 0.94 mas). In order to
derive the orbital parameters, we requested all the observations
of DG~Leo~AB from the Washington Double Star Observations Catalog
\citep{2001AJ....122.3466M} and obtained 44 micrometric and 38
speckle-interferometric measurements (see Table~\ref{tab:vor}).
We further received one unpublished speckle observation
(Hartkopf, W., 2004, private comm.). Since 1997 the systematic
observations were discontinued owing to the difficulty of access
to sufficiently large telescopes for this type of object.
Unfortunately, the astrometric data in themselves are
insufficient to compute a reliable visual orbit, as practically no
change in position angle was detected during a time interval of
more than 40 years.
To complete the data we therefore included the components' radial
velocities
(see Table~\ref{tab:vrc}) for a combined
astrometric--spectroscopic orbit analysis. Use was made of the
software package developed by \citet{1998A&AS..131..377P}. It is
based on the principles of simulated annealing \citep[SA,
][]{metropolis} for the global exploration and minimization in
the parameter space followed by a local least-squares
minimization (following the scheme of
Broyden-Fletcher-Goldstrab-Shanno). Table~\ref{tab:vop} lists the
stable and well-defined orbital parameters corresponding to one
of the best solutions in the sense of least-squared
residuals after 1500 iterations with SA and exploring the period
range 70--200 years. Fig.~\ref{fig:speckle} illustrates this
orbit, both for the visual {(upper panel)} and the spectroscopic
{(lower panel)} datasets (cf. Table~\ref{tab:vrc}). The accuracy
and the confidence level of this solution are further discussed
in Sect.~\ref{sec:dvb}.

\begin{table}
\center
\begin{minipage}{7.cm}
\center \caption{Radial velocities of components A and B. FB77:
\citet{1977PASP...89..216F}; RS91: \citet{1991PASP..103..628R};
EB2002: ELODIE observations made by \citet{O02}; PN2003: ELODIE
observations performed by \citet{N03}; PM2004: ELODIE observations
obtained by \citet{M04}.} \label{tab:vrc}
\begin{tabular}{@{}lrrl@{}}
\hline JD & V{\sc a} & V{\sc b} & Ref. \\
& (\kps) & (\kps)&   \\ \hline
2440995.9042 & 26.9 $\pm$ 3  & 28.1 $\pm$ 5 & FB77\\
2442113.9789 & 25.1 $\pm$ 2  & 27.9 $\pm$ 4 & FB77\\
2442470.8851 & 27.2 $\pm$ 3  & 29.5 $\pm$ 3 & FB77\\
2442856.8086 & 29.3 $\pm$ 2  & 24.3 $\pm$ 4 & FB77\\
2447815.7210 & 29.9 $\pm$ 2  & 22.2 $\pm$ 2 & RS91\\
2452300.6250 & 19.3 $\pm$ 2  & 41.6 $\pm$ 4 & EB2002\\
2452649.0512 & 20.1 $\pm$ 1 & 38.5 $\pm$ 4 & this work\\
2452771.3452 & 21.1 $\pm$ 4  & 43.0 $\pm$ 5 & PN2003\\
2453026.6498 & 22.0 $\pm$ 4  & 32.0 $\pm$ 5 & PM2004\\
\hline
\end{tabular}
\end{minipage}
\end{table}


\setlength{\tabcolsep}{0.8mm}
\begin{table}
\center
\begin{minipage}{7cm}
\caption[]{\label{tab:vor} Sample of micrometric ($\theta$,
$\rho$) and speckle (x, y). The data are followed by the residuals
(O$-$C). A full version of this table is available at the CDS.}
\center
\begin{tabular}{crrrr}
 \hline
\multicolumn{5}{c}{\sc micrometric data}\\
\hline
  Epoch &
  $\theta$(\degr)&(O$-$C)$\theta$&$\rho$(\arcsec)&(O$-$C)$\rho$\\
  \hline
 1935.4400& 218.360&   $+$3.548&  0.335&  $-$0.005\\
 1936.8700& 218.118&   $-$5.715&  0.339&  $-$0.019\\
 1937.1100& 218.078&  $+$11.524&  0.339&  $-$0.039\\
 1937.2800& 218.049&   $+$8.852&  0.340&  $-$0.040\\
 1938.1600& 217.904&   $+$0.195&  0.342&  $-$0.042\\
 ...\\
  \hline
\multicolumn{5}{c}{\sc speckle data}\\
\hline
 Epoch&x(")&(O$-$C)x&y(")&(O$-$C)y \\
 \hline
 1940.1200& $-$0.274& $-$0.010& $-$0.211& $+$0.008\\
 1976.8606& $-$0.249& $-$0.001& $-$0.152& $-$0.004\\
 1977.1802& $-$0.247& $+$0.011& $-$0.151& $-$0.032\\
 1977.3331& $-$0.247& $+$0.002& $-$0.150& $-$0.006\\
 1978.3161& $-$0.242& $-$0.015& $-$0.146& $+$0.001\\
 ...\\
 \hline
\end{tabular}
\end{minipage}
\end{table}

\begin{figure}
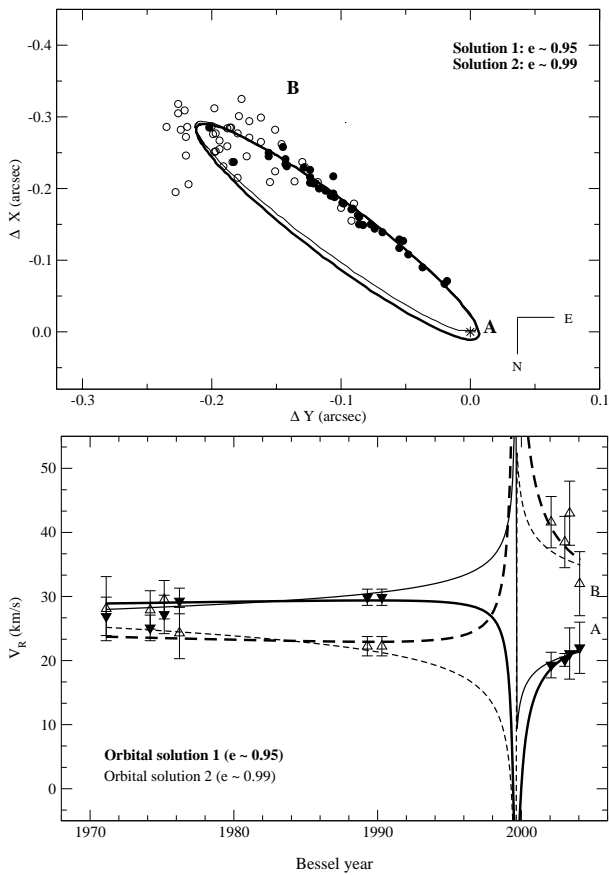
\center
\begin{tabular}{c}
\includegraphics[width=8cm,angle=0,clip=]{dgvmer1.eps}\\
\includegraphics[width=8cm,angle=0,clip=]{radz.eps}
\end{tabular}

\caption{{\bf Upper panel:} Observations and computed relative
visual orbits: the best solution is represented by the thick
solid curve while the other represents an orbit with an
eccentricity very close to 1. The micrometric and speckle data
obtained from 1935 to 1997 are represented by open and filled
circles respectively. {\bf Lower panel:} Observed (comp A: open
triangles -- comp B: filled triangles) and predicted (comp A:
solid curve and comp B: dashes) radial-velocity variations for
the two solutions (thick and thin curves) shown in the upper
panel (see discussion in Sect.~\ref{sec:dvb}).}
\label{fig:speckle}
\end{figure}

\begin{table}
\center
\begin{minipage}{7cm}
\centering \caption[]{\label{tab:vop} Values and standard
deviations of the constrained parameters of the combined
astrometric-spectroscopic orbital solution of DG Leo AB.}
\begin{tabular}{lr@{}l@{}lr@{}l@{}l}
\hline
Orbital parameter &\multicolumn{3}{c}{Value} & \multicolumn{3}{c}{Std. dev.}\\
\hline
$a$ (\arcsec)&0&.&191 &0&.&015\\
$i$ (\degr)&117&.& &13&.& \\
$\omega$ (\degr)&341&.& &13&.&\\
$\Omega$ (\degr)&27&.&2 &3&.&0\\
$e$&0&.&946 &0&.&043\\
$P$ (yr)&102&.&3 &9&.&5\\
$T$ (Besselian year)&1999&.&59 &0&.&41\\
$V_0$ (\kps)&+26&.&97 &0&.&49\\
$\kappa=\frac{M_{\rm B}}{M_{\rm A}+M_{\rm B}}$&0&.&374 &0&.&046\\
\hline
\end{tabular}
\end{minipage}
\end{table}

\section{Atmospheric fundamental parameters}

\label{sec:fundpar}

{The disentangled component spectra provided by {\sc korel} allow
the use of well known techniques of spectral analysis that lead
to the determination of physical quantities (i.e.: \teff,
\logg~and \vsini) accurately describing the stellar atmospheres.}
For DG Leo's components, these atmospheric fundamental parameters
were derived in three consecutive steps which are chronologically
described in the following subsections and the results of which
are listed in Table~\ref{tab:fund}.

\subsection{Projected rotation velocity}

{A Fourier analysis of several unblended line profiles was used
to determine the projected rotation velocity (V~sin~{\it i}). The
adopted procedure that uses a FFT {\sc fortran} subroutine
provided by \citet{1992oasp.book.....G}} is similar to the one
described by \citet{1933MNRAS..93..478C} and largely applied for
example by \citet{2002A&A...381..105R}. It is based on the
location of the first minimum of the Fourier transform that
varies as the inverse of V~sin~{\it i}. Twelve isolated spectral
lines were therefore chosen in the disentangled spectra (see
Sect.~4). The projected rotation velocities thus derived are
given in Table~\ref{tab:fund}. The listed error is equal to the
root mean square of the measurements made on each of the 12
unblended lines. Our \vsini~measurements confirm the general
trend that is observed when the widths of the cross-correlation
functions (Fig.~\ref{fig:crossc}) for each component are compared
to each other: DG~Leo~Aa is apparently rotating faster than
DG~Leo~Ab. This unexpected result is discussed in Sect.~8.5.

\begin{figure*}
\center
\includegraphics[width=16cm,angle=0,clip=]{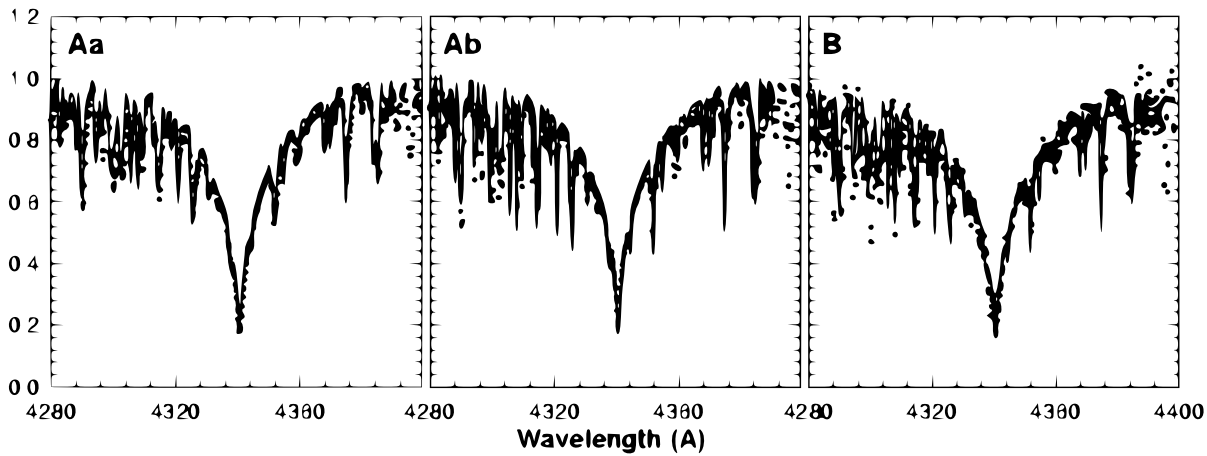}
\caption{Fitting of the observed hydrogen line profiles (dots)
with theoretical spectra (line).} \label{fig:fundhg}
\end{figure*}

\subsection{Effective temperature}

\label{subsec:teff}

In the atmospheres of mid- to late-A-type stars, hydrogen exists
mostly in a neutral form. Hydrogen lines are therefore rather
insensitive to surface gravity and can be used to derive the
effective temperature \citep[see for
example:~][]{1993pvnp.conf..182S}. Consequently, we estimated the
effective temperature of DG Leo's components by fitting the
individual H$\alpha$ and H$\gamma$ lines with theoretical line
profiles (Fig.~\ref{fig:fundhg}). {Fully line-blanketed model
atmospheres were computed in LTE with {\sc atlas 9}
\citep{cdrom13} corrected to account for the comments made by
\citet{1997A&A...318..841C}.} From these models, a flux grid was
obtained by use of the {\sc synspec} computer code \citep[][ see
references therein]{1995ApJ...439..875H} enabling the IRSCT and
IOPHMI opacity flags, respectively, to account for the Rayleigh
scattering and for the H$^{-}$ ions. A least-squares method and
the {\sc minuit} minimization package of CERN were finally used to
fit the observed line profiles. During the procedure, the surface
gravity was kept fixed to 4. (cgs) for all three components.
Error bars were computed from the largest deviation obtained by
performing the fit several times and assuming different start
values for the effective temperature.

\subsection{Surface gravity}

{Because in late-A-type stars the hydrogen lines are insensitive
to a change in surface gravity, we followed the same procedure as
used by \citet{1997hipp.conf..239N} and
\citet{2003A&A...398.1121E} for stars within a distance of 150 pc
to derive the surface gravity of the components. The luminosity
was calculated from the {\sc hipparcos} parallax of the triple
star and the components' V magnitudes (see Table~\ref{tab:lr}).
The bolometric correction was adopted from
\citet{1996ApJ...469..355F}. Making use of the bolometric
magnitude and the previously derived effective temperature, we
then obtained the mass and the radius of each component through
interpolation in the theoretical evolutionary tracks of
\citet{1992A&AS...96..269S} (for Z=0.001 and 0.020). In this way
the surface gravities of the components were found to be much
alike: the values are listed in Table~\ref{tab:fund} together
with the interpolated masses (M$_{\rm{HR}}$) and radii
(R$_{\rm{HR}}$). Their accuracy was estimated from the errors
affecting the {\sc hipparcos} parallax
($\frac{\sigma_\pi}{\pi}$=15\%) and the effective temperature.}

\begin{table}
\center \caption{Derived fundamental parameters} \label{tab:fund}
\begin{tabular}{@{}lr@{}c@{}lr@{}c@{}lr@{}c@{}l@{}}
\hline {Star} & \multicolumn{3}{c}{\bf Aa} &
\multicolumn{3}{c}{\bf Ab} & \multicolumn{3}{c}{\bf B} \\ \hline
{T$_{\rm{eff}}$} (K) & 7470 &$\pm$& 220        & 7390 &$\pm$& 220           & 7590 &$\pm$& 220  \\
{log$~g$} & 3.8 &$\pm$& 0.14         & 3.8 &$\pm$& 0.14           & 3.8 &$\pm$& 0.12 \\
{V~sin~{\it i~}}  ($\rm{km~s^{-1}}$)            & 42 &$\pm$& 2  & 28 &$\pm$& 2 &  31 &$\pm$& 3  \\
{$\xi_{\rm{turb}}$} ($\rm{km~s^{-1}}$) &   2.3 &$\pm$& 0.5   & 2.3 &$\pm$& 0.5  &  2.5 &$\pm$& 0.5 \\
M$_{\rm{HR}}$ (M$_\odot$) &  2.0 &$\pm$& 0.2      & 2.0 &$\pm$& 0.2      &  2.1 &$\pm$& 0.2    \\
R$_{\rm{HR}}$ (R$_\odot$) &  2.96 &$\pm$& 0.20 &  2.94 &$\pm$&
0.20 &
2.99 &$\pm$& 0.20\\
\hline
\end{tabular}
\end{table}

\section{Abundance analysis}

\label{sec:abundance analysis}

The disentangled spectra were used to estimate the components'
atmospheric chemical composition. We therefore systematically
applied {\sc korel}, order by order and from 4900 \AA~to 6100
\AA, keeping the orbital parameters and luminosity ratios fixed
to the values in Table \ref{tab:lr} and Sect.~\ref{subsec:OPSB}.
The abundance determination was then carried out by fitting
theoretical line profiles to the observed components' spectra (see
Fig. \ref{fig:fit}).

\subsection{Atomic data}

Oscillator strengths, energy levels and damping parameters
(including Stark, van der Waals and natural broadening) used
during the fitting procedure were basically those compiled in the
VALD-2 database \citep{1999A&AS..138..119K} updated by
\citet{2002A&A...383..227E}. For elements having atomic numbers
greater than 30, the line list was completed by atomic data
extracted from Kurucz CD-ROM N23 \citep{cdrom23} and from the NIST
Atomic Spectra Database \citep[see the gfVIS.dat file created
by][]{tlustyh}. \ion{Sc}{ii} and \ion{Mn}{ii} were treated {with}
oscillator strengths proposed by \citet{2000A&A...353..722N}.

\subsection{Chemical composition}

\begin{figure}
\center
\includegraphics[width=8.5cm,angle=0,clip=]{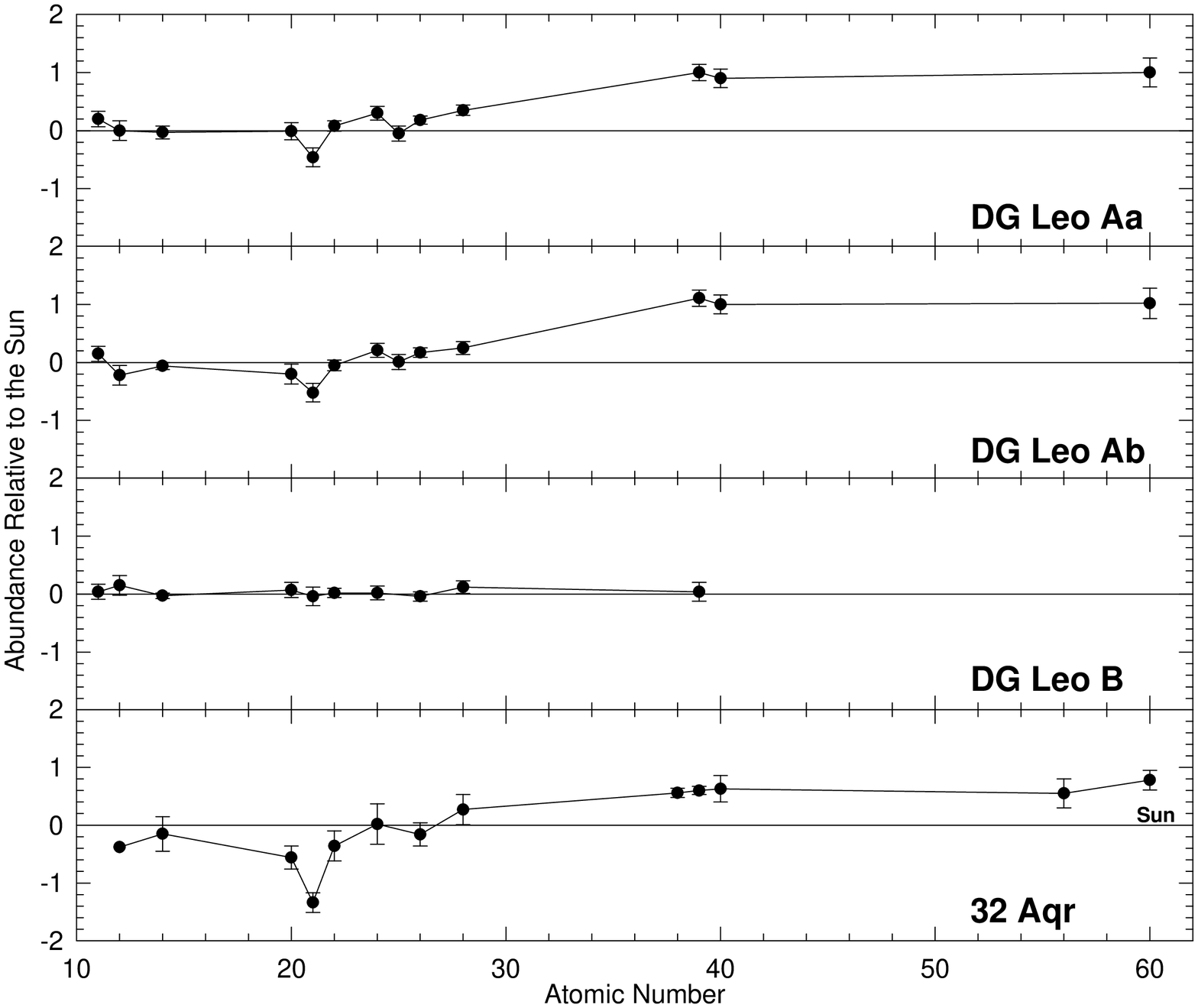}
\caption{The abundance patterns we observed for the components of
DG Leo are compared to those determined in the Sun
\citep{1998SSRv...85..161G} and in the Am star 32 Aqr
\citep{1993pvnp.conf..213K}.} \label{fig:pattern}
\end{figure}

The microturbulent velocity was supposed to be constant with
optical depth. It was derived simultaneously with the iron
abundance by fitting the observations, from 4950 \AA~to 6100 \AA,
with theoretical spectra. Except for iron, during this procedure
the chemical composition was kept fixed at the solar values. The
resulting microturbulent velocities, $\xi_{\rm{turb}}$, are given
in Table \ref{tab:fund} and were further used to derive the
abundances of the other atomic species.

A description of the model atmospheres and of the computer code
we used can be found in Sect.~\ref{subsec:teff}. For the same line
lists and model atmospheres, there is generally a good agreement
between the synthetic spectra provided by {\sc synspec46} and
those computed by other programs such as {\sc spectrum} which was
written by R.O. Gray \citep[e.g.:][]{1994AJ....107..742G}.
However, \ion{Sc}{ii} line profiles were systematically much
stronger in {\sc synspec}'s spectra, providing scandium
abundances one order of magnitude smaller than expected for our
reference star. The use of Kurucz's {\sc pfsaha} subroutine, to
compute the partition functions of scandium, definitely resolved
this incoherence. It is worth noting that this problem does not
occur with the older versions of {\sc synspec}.

\setlength{\tabcolsep}{0.9mm}
\begin{table}
\caption{Components' chemical compositions relative to hydrogen}
\label{tab:abundances} \center
\begin{tabular}{@{ }lcccrr@{ }}
\hline
                 &   \multicolumn{3}{c}{\bf DG Leo} &   {\bf n}               &  {\bf Sun}     \\
                & $\epsilon$(Aa)             & $\epsilon$(Ab)          & $\epsilon$(B)          &           &
                \\\hline
  Na            & $-$5.47 $\pm$ 0.13  & $-$5.52 $\pm$ 0.13  & $-$5.63 $\pm$ 0.13  & 1  &  $-$5.67 \\
  Mg            & $-$4.42 $\pm$ 0.17  & $-$4.64 $\pm$ 0.17  & $-$4.27 $\pm$ 0.17  & 2  &  $-$4.42 \\
  Si            & $-$4.48 $\pm$ 0.11  & $-$4.51 $\pm$ 0.06  & $-$4.48 $\pm$ 0.05  & 3  &  $-$4.45 \\
  Ca            & $-$5.65 $\pm$ 0.15  & $-$5.84 $\pm$ 0.17  & $-$5.57 $\pm$ 0.13  & 4  &  $-$5.64 \\
  Sc            & $-$9.29 $\pm$ 0.16  & $-$9.35 $\pm$ 0.16  & $-$8.87 $\pm$ 0.16  & 2  &  $-$8.83 \\
  Ti            & $-$6.74 $\pm$ 0.09  & $-$6.87 $\pm$ 0.09  & $-$6.80 $\pm$ 0.08  & 5  &  $-$6.82 \\
  Cr            & $-$6.03 $\pm$ 0.12  & $-$6.12 $\pm$ 0.12  & $-$6.31 $\pm$ 0.12  & 8  &  $-$6.33 \\
  Mn            & $-$6.66 $\pm$ 0.13  & $-$6.60 $\pm$ 0.13  &                   & 2  &  $-$6.61 \\
  Fe            & $-$4.32 $\pm$ 0.07  & $-$4.33 $\pm$ 0.08  & $-$4.54 $\pm$ 0.08  & 30 &  $-$4.50 \\
  Ni            & $-$5.40 $\pm$ 0.09  & $-$5.50 $\pm$ 0.11  & $-$5.63 $\pm$ 0.11  & 7  &  $-$5.75 \\
  Y             & $-$8.78 $\pm$ 0.14  & $-$8.67 $\pm$ 0.14  & $-$9.74 $\pm$ 0.16  &  5 &  $-$9.78 \\
  Zr            & $-$8.49 $\pm$ 0.16  & $-$8.39 $\pm$ 0.16  &                   &  1 &  $-$9.39 \\
  Nd            & $-$9.50 $\pm$ 0.26 &  $-$9.48 $\pm$ 0.26 &                   &  4 &  $-$10.50 \\
\hline
\end{tabular}
\end{table}

The final results of the chemical analysis are summarized in Table
\ref{tab:abundances} and in Fig.~\ref{fig:pattern}. {An example
of the agreement obtained between the fitted synthetic spectra
and the disentangled ones is shown in Fig.~\ref{fig:fit}.}
Logarithmic abundance values are given relatively to hydrogen for
the Sun (Col. 6) as well as for each component of DG Leo (Col. 2,
3 and 4). Solar abundances were taken from
\citet{1998SSRv...85..161G}. The number of transitions considered
to derive the atmospheric chemical composition of DG Leo's
components is noted in Col. 5. Abundance error bars,
$\Delta\epsilon$, were computed {in the framework of
non--correlated error sources with the following relation:}

\begin{equation}
\sigma^2~=~{{\sigma^2_{\rm{scat}}} \over
{n-1}}~+~\Delta\epsilon{\rm{(Teff)}}^2~+~\Delta\epsilon{\rm{(log~g)}}^2~+~\Delta\epsilon{\rm
(\xi_{\rm{turb}})}^2
\end{equation}

\noindent where n is the number of lines that we used and where
$\sigma_{\rm{scat}}$, $\Delta\epsilon{\rm{(Teff)}}$,
$\Delta\epsilon{\rm{(log~g)}}$ and $\Delta\epsilon{\rm
(\xi_{\rm{turb}})}$ are the errors related respectively to the
r.m.s. scatter of the derived abundance, to the effective
temperature, surface gravity and microturbulent velocity accuracy.
The uncertainties $\Delta\epsilon$ on the abundance introduced by
the fundamental parameters are given in Table~\ref{tab:sigma}.
They were computed relatively to a reference atmosphere model
defined by the following fundamental parameters: \teff~=~7500~K,
\logg~=~4.0, $\xi_{\rm{turb}}$ = 2.5 \kps~and solar chemical
composition. When only one line is available to determine the
abundance, the computation of $\sigma_{\rm{scat}}$ is based on
the oscillator strength's relative error (fixed at 25\% when not
found in the NIST database).

\begin{figure*}
\center
\begin{tabular}{c}
\includegraphics[width=11cm,angle=270,clip=]{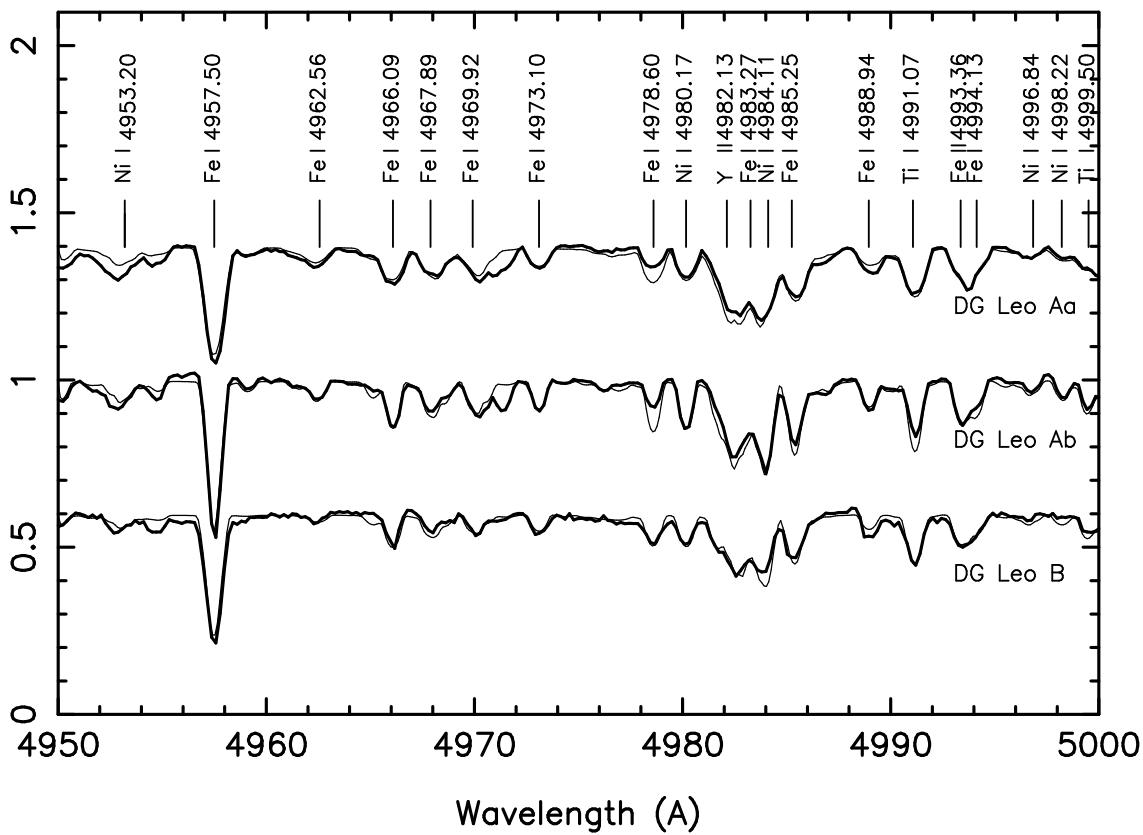}\\
\includegraphics[width=11cm,angle=270,clip=]{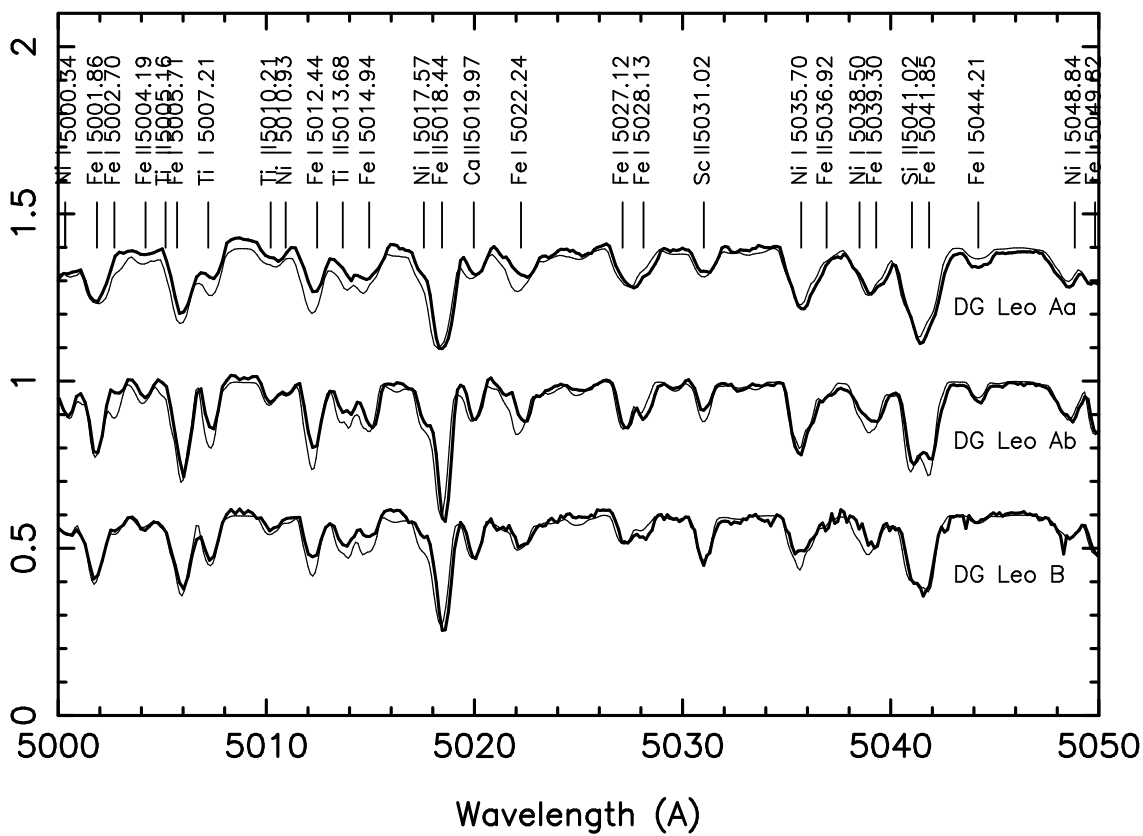}
\end{tabular}
\caption{Comparison between the vertically shifted normalized
component spectra (thick line) and the fitted synthetic spectra
resulting from the chemical abundance analysis (thin line).
Identifications indicate the transitions contributing the most to
the spectral lines.} \label{fig:fit}
\end{figure*}

\begin{table}
\center \caption{Abundance deviations ($\Delta\epsilon$) caused
by the upper and lower limits of \teff, \logg~and microturbulent
velocity ($\xi_{\rm{turb}}$). }\label{tab:sigma}
\begin{tabular}{lccc}
\hline
 Elem.        & $\Delta\epsilon{\rm{(Teff)}}$ & $\Delta\epsilon{\rm{(log~g)}}$  &   $\Delta\epsilon{\rm (\xi_{\rm{turb}})}$
 \\\hline
  Na            &  0.03 & 0.02 & 0.05 \\
  Mg            &  0.10 & 0.04 & 0.07 \\
  Si            &  0.02 & 0.02 & 0.00 \\
  Ca            &  0.03 & 0.01 & 0.06 \\
  Sc            &  0.01 & 0.06 & 0.13 \\
  Ti            &  0.02 & 0.05 & 0.05 \\
  Cr            &  0.06 & 0.01 & 0.03 \\
  Mn            &  0.05 & 0.00 & 0.04 \\
  Fe            &  0.01 & 0.00 & 0.06 \\
  Ni            &  0.02 & 0.03 & 0.05 \\
  Y             &  0.02 & 0.05 & 0.09 \\
  Zr            &  0.02 & 0.11 & 0.05 \\
  Nd            &  0.08 & 0.07 & 0.05 \\
\hline
\end{tabular}
\end{table}

\section{Discussion}

\subsection{Spectroscopic orbit of the close binary Aab}

\label{sec:dsb}

The parameters we derived for the spectroscopic binary are in good
agreement with the previously published values
\citep{1977PASP...89..216F,1991PASP..103..628R}. From the
{spectroscopic} orbital parameters {listed in Sect 5.1}, one may
further {derive}:

\begin{eqnarray}
(\rm{a_{Aa}+a_{Ab}})\sin i_{\rm A} &=& \rm{16.5}\pm\rm{0.02~R_{\odot}}\label{eq:a}\\
\rm{M_{Aa}}\sin^3 i_{\rm A} &=& \rm{1.760}\pm\rm{0.007~M_{\odot}}\label{eq:b}\\
                    &=& \rm{M_{Ab}}\sin^3 i_{\rm A}\label{eq:c}
\end{eqnarray}

\noindent where $i_{\rm A}$ stands for the inclination of the
close binary and {where we introduced} the values of K, q, P and
e as well as their associated error bars. Since no eclipses
{were} observed \citep{phot1}, the derived orbital parameters
{constrain the orbital inclination angle} which will depend on
the stellar radius and thus on the surface gravity. {From the
condition $\rm{R_{Aa}} +\rm{R_{Ab}} < (\rm{a_{Aa}+a_{Ab}}) \cos
i$ (no eclipses), we have that:}
\begin{center}
\begin{tabular}{rcl}
$\tan~i_{\rm A}$ &$<$& ${(\rm{a_{Aa}+a_{Ab}})~\sin~i_{\rm A}}
\over {R_{Aa}+R_{Ab}}$
\end{tabular}
\end{center}

For $\rm{R_{Aa}}=\rm{R_{Ab}}=R$ and {if we use}
$\rm{R^{\rm{-1}}}\propto(g/M)^{1/2}$ in relations (\ref{eq:a})
and (\ref{eq:b}), we obtain:

\begin{eqnarray}
\sin i_{\rm A} \cos^2i_{\rm A} &>& \frac{{\rm M}\sin^3i_{\rm
A}/\rm{M_{\odot}}}{\left( {\rm{(a_{Aa}+a_{Ab})}}/{2} \right)^{2}}
\frac{\rm g_\odot}{\rm
g}\\
&>& 0.02586 ~\frac{\rm g_\odot}{\rm g}
\end{eqnarray}

\noindent Numerically for $\log g=3.8\pm0.14$ (see
Table~\ref{tab:fund}), we find:

\begin{equation}
i_{\rm A}< 69^{\rm o}_{\rm .}8^{\rm+3.2}_{\rm-4.2}
\label{eq:angle}
\end{equation}

\noindent implying an upper limit for the orbital inclination of
73$^{\rm o}$ from the absence of eclipses. Comparing the stellar
masses derived from the evolution tracks (see M$_{\rm{HR}}$ in
Table~\ref{tab:fund}), it is clear that the orbital inclination
cannot be much lower, as the lower mass limit deduced from
(\ref{eq:angle}) and (\ref{eq:c}) gives:

\begin{eqnarray}
\rm{M_{Aa}}=\rm{M_{Ab}}&>&2.1^{+0.2}_{-0.1}~\rm{M_{\odot}}.
\end{eqnarray}

\subsection{Astrometric--spectroscopic orbit of the wide binary AB}

\label{sec:dvb}

Recent and old astrometric data were combined with radial
velocities obtained at an epoch close to periastron passage.
Using different starting points corresponding to the 100 lowest
values of the objective function found with SA, we ended up with
the solution proposed in Sect. 5.2 in 80\% of the cases after a
fine tuning of the minimization process. In only 3\% of the
cases, SA converged successfully to solutions with some parameter
out of the 1.5-$\sigma$ uncertainties given in
Table~\ref{tab:vop}. Although the residuals are realistic given
the measurement errors, the orbit is {\em not} yet definitive.
Some parameters remain unconstrained because we miss radial
velocities near periastron in a very eccentric orbit. At present,
we conclude that:
\begin{itemize}
\item[--] the orbital period is notably shorter than previously estimated
by \citet{1977PASP...89..216F}, with a most probable value close
to 100 years (and an uncertainty of 10\%);
\item[--] the orbit is definitely very eccentric, but unfortunately
we do not have radial velocities near periastron. Any spectra
taken in the interval 1998--2002 would be of immense value if
existing;
\item[--] the fractional mass is constrained reasonably well with a most
probable value $\kappa$ = 0.37, and consistent with the value
derived from the fundamental atmospheric parameters ($\kappa$ =
0.34 $\pm$ 0.07). The constraint on the orbital $\kappa$ already
induces a useful lower limit to the mass of component B relative
to each of the equal-mass components of A:
\begin{eqnarray}
\frac{\rm{M_{B}}}{\rm{M_{A_{a}}}} &>& 1.19^{-0.21}_{+0.26}.
\end{eqnarray}
In other words, component B is very probably at least as massive
as each of the close binary components;
\item[--] the other constrained parameters (Table~\ref{tab:vop}) include the orbital inclination
(range 110--130 degrees) and those related to the astrometric
data;
\item[--] ambiguity remains with several parameters related to the unconstrained
radial-velocity amplitude (see the two different solutions shown
in Fig. 3). As a consequence, we do not know the true semi-axis
major (in \kps), which would provide a dynamical parallax and
absolute masses.
\end{itemize}

Hence, we can use the mass of component A obtained from the close
spectroscopic orbit. From Table~\ref{tab:fund} and eq. 8, we have
$M_A = 4.2 \pm 0.2~M_{\sun}$. Then the data of
Table~\ref{tab:vop}, together with standard--error propagation
laws, lead to:
\begin{eqnarray}
\rm{M_{B}}&=&2.5\pm0.6~\rm{M_\odot}\nonumber\\
\rm{K_A}&=&12.3\pm5.8~\rm{km~s}^{\rm{-1}}\nonumber\\
\rm{K_B}&=&20.6\pm8.7~\rm{km~s}^{\rm{-1}}\nonumber\\
\rm{a}=\rm{a_A+a_B}&=& 41.2\pm5.8~\rm{AU}\nonumber\\
\pi&=&4.6\pm1.0~\rm{mas}, \nonumber
\end{eqnarray}

\noindent where the dynamical parallax is consistent with the
Hipparcos parallax at the 1.3--$\sigma$ level.

A next step would be to repeat the SA procedure limiting the
exploration of the parameter space to realistic values of
$\rm{M_{A}}$. However, significant progress in the direct
determination of the individual masses with a sufficiently high
accuracy could be achieved, either by including possibly existing
radial velocities near periastron passage -- we therefore invite
anyone who might have spectra or radial velocities of DG Leo
during the interval 1998--2002 (see Fig.~\ref{fig:speckle}) to
contact us -- or by gathering new speckle-interferometric
observations in the following years.

\subsection{Ca abundance and luminosity ratios}

\label{sec:calr}

As noted in Sect.~\ref{sec:korel}, the luminosity ratios of the
three components were derived {under the assumption} that their
Ca~{\sc ii}~K line-depths are saturated and identical. To test the
validity range of that assumption, we compare in Fig. \ref{fig:ca}
the theoretical spectra of all three components of DG Leo from
3920 to 3950 \AA. Owing to the close values of the fundamental
parameters and to the fact that the calcium underabundance
observed in the close binary components is smaller than 0.20 dex,
the line depths are very similar. As a matter of fact, the
slightly smaller depth in the spectrum of the faster rotating
component Aa might have introduced a flux zero--point error
smaller than 1.5~percent, which does not significantly affect the
abundance analysis.

\begin{figure}
\center
\includegraphics[width=7.5cm,angle=0,clip=]{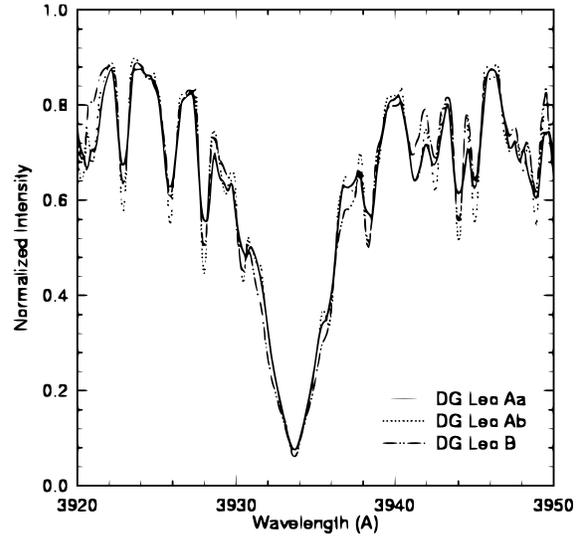}
\caption{Computed Ca~{\sc ii}~K line-profile for all three
components of DG Leo.} \label{fig:ca}
\end{figure}

\begin{figure}
\center
\includegraphics[width=8.5cm,angle=0,clip=]{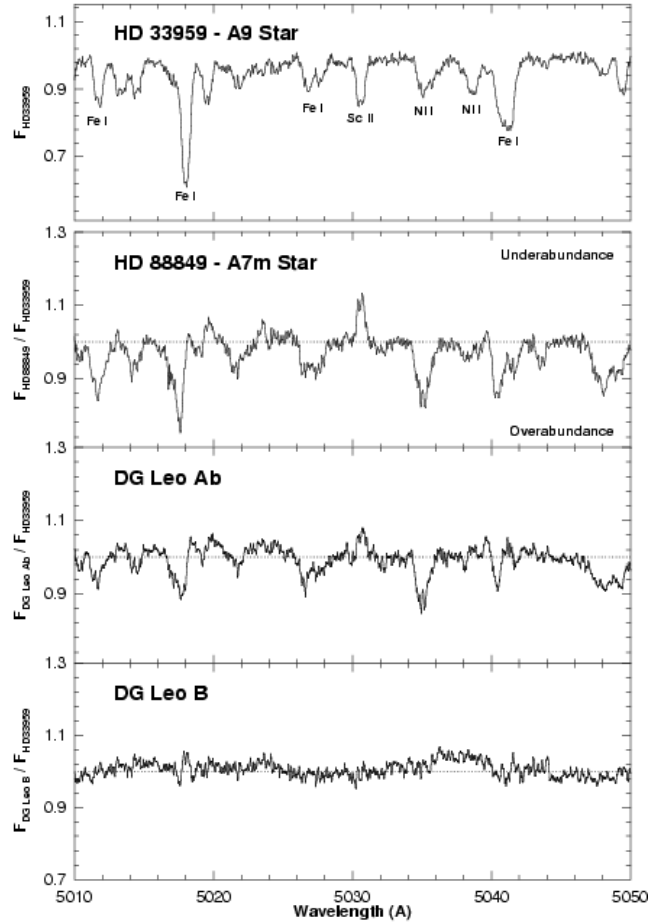}
\caption{The spectrum of the chemically normal star HD 33959
(top) and the ratio of the Am and DG Leo spectra (components Ab
and B only) to the top one are plotted.} \label{fig:comp}
\end{figure}

\subsection{Comparison with reference stars}

{If we discount the larger effects of Doppler broadening} for DG
Leo Aa, both components of the close binary have very similar
spectra. Their nature is however different from that of component
B and their iron-peak lines are generally significantly stronger.
To visualize the general trend that can be noted in the chemical
abundances we derived, their respective spectra have been
compared to the spectra of the A9 star HD~33959 (14~Aur) and the
metal-line A7m star HD~88849 (HR~4021) found in {\sc The Elodie
Archive} at OHP. These stars are similar to DG~Leo's components
in spectral type and projected rotation velocity, but different
in chemical composition. To make this more obvious, the spectra
of HD~88849 and those of the Ab and B components were divided by
the reference spectrum of HD~33959 and reported in Fig.
\ref{fig:comp}. According to this comparison, DG Leo B is found
to have almost the same chemical composition and fundamental
parameters as the reference star which is known to be {\it
chemically normal} \citep{2000A&AS..144..203H}. For their part,
the A components present typical Am-type peculiarities (e.g.:
scandium deficiency and enhanced iron-peak elements) but somewhat
weaker than those observed in HD~88849. {From a} quantitative
point of view, our abundance analysis clearly shows that in the A
components' atmospheres, iron-peak elements are overabundant by up
to 0.2 dex relatively to the Sun and by up to 1.0 dex for the
heavier elements (Y, Zr, Nd). The overabundances are therefore of
the same order of magnitude as is observed in other typical Am
stars such as 32 Aqr (HD~209625, see Fig. \ref{fig:pattern}).
Deficient species, especially calcium, are less extremely
underabundant in the close binary components of DG Leo.

\begin{figure}
\center
\begin{tabular}{cc}
\includegraphics[width=6cm,angle=270,clip=]{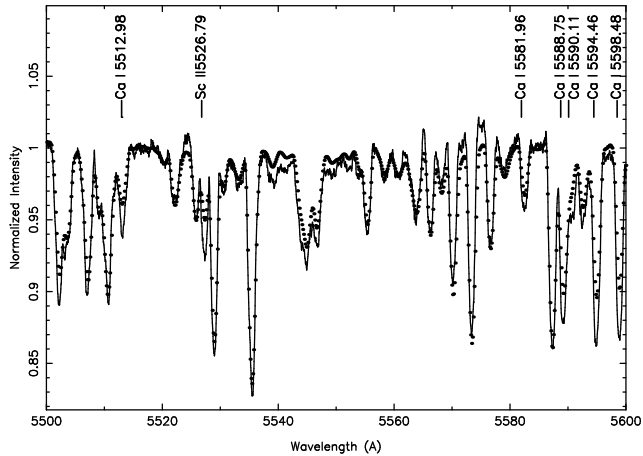}
\end{tabular}
\caption{The spectra of DG Leo Aa (line) and DG Leo Ab (dots) are
compared. The spectrum of the latter component was convolved with
a Gaussian function in order to {increase artificially} its
projected rotation velocity.} \label{fig:difA}
\end{figure}

\subsection{The close binary components Aa and Ab}

Spin--orbit synchronization is generally expected to occur 100
times faster than orbital circularization
\citep{1977A&A....57..383Z}. Stars in an already circularized
orbit are therefore expected to be synchronized and to have
similar projected rotation velocities. As mentioned earlier, this
is obviously not the case for the close binary of DG Leo. We note
indeed (Sect.~6.1) a significant disagreement between the
\vsini~values of the two components. The predicted synchronization
velocity V$_{\rm synch}$=36$\pm$6~\kps~lies in between the
measured values. Therefore, we cannot exclude that one of the
components would be synchronized. This certainly merits further
attention.

Components of multiple systems such as DG Leo {usually} originate
from the same protostellar environment. Their similarity or
difference in chemical surface composition therefore relates
{directly} to their {individual/uncoupled} evolution after
formation (stellar evolution, rotation, diffusion), and the
possible influence of a close companion. It is known from
Sect.~\ref{sec:fundpar} that the Aa and Ab components have similar
fundamental parameters. As can be noted from
Table~\ref{tab:abundances} and from Fig. \ref{fig:pattern}, they
further have very similar abundance patterns. Both components are
Am stars showing the same {magnitude of} overabundance for the
iron-peak elements {as well as} for the heavier ones. {The
underabundances are mild compared to many classical Am stars. A
differential comparison of the disentangled spectra of Aa and Ab
(Fig.~\ref{fig:difA}) gives evidence for a marginally higher Ca
abundance in Aa. This fact remains unnoticed from
Table~\ref{tab:abundances} because the uncertainties given on the
abundances include the uncertainty on the fundamental atmospheric
parameters at an {\it absolute} level. When comparing Aa and Ab,
only the {\it relative} errors in $\Delta$\teff, $\Delta$\logg~and
$\Delta\xi_{\rm{turb}}$ are relevant.}


Classical diffusion models \citep{1996A&A...310..872A} generally
show that calcium and scandium are leaving the atmosphere after a
while. Observations
\citep{1993pvnp.conf..577N,1998A&A...330..651K} and numerical
simulations \citep{1996A&A...310..872A} predict a short
phase of calcium overabundance at the beginning of the main
sequence followed by a calcium abundance decrease. In this sense,
abundances are suspected to be very sensitive to mass loss,
rotation and age which may explain the somewhat different
behaviour regarding diffusion observed in the {components Aa and
Ab of DG Leo}.

}



\subsection{Spectral line variability}

\label{sec:pul}

\begin{figure}
\center
\includegraphics[width=8cm,angle=0,clip=]{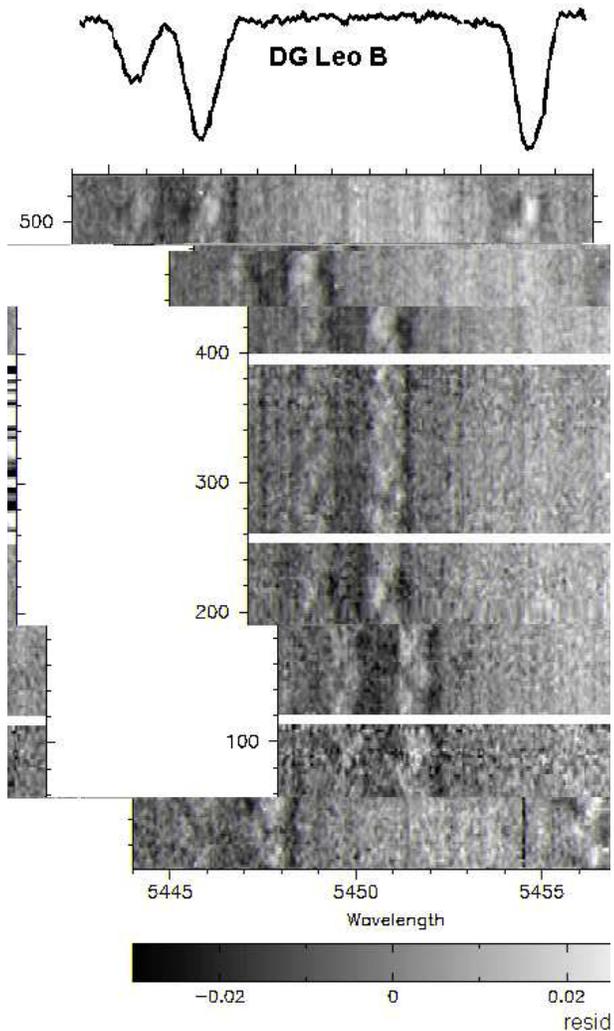}
\caption{The residuals provided by {\sc korel} are plotted in a
gray scaled figure for 4 consecutive nights (January 11 -- 15,
2003). Each variation is centred on the spectrum of component B
and repeats with a period of about 2 hours.} \label{fig:resid}
\end{figure}

The residuals of {\sc korel} clearly indicate that the
spectroscopic variations due to pulsations are centred on the
spectrum of component B (Fig.~\ref{fig:resid}). The pattern of
these variations is reproduced with a regularity of about 2~hours
which agrees with the results from the photometry \citep{phot1}.
A detailed study of the line variability awaits the analysis of
the photometry as both sources should be combined in a
periodicity analysis (our time base for spectroscopy is very
short) to enhance the probability of identifying the pulsation
modes. However, we note already the multi-periodic character of
the line variability from the lack of exact repetition of the
variability patterns in different nights.

Despite this line variability, we further notice that the
time-averaged spectrum of each component of DG Leo is fully
consistent with the combined observed spectra and no numerical
artifacts were detected. The use of {\sc korel} to disentangle
the component spectra from the high-resolution data may thus be
considered as highly successful in this specific case.

\section{Conclusions}

The Fourier Transform spectral disentangling technique developed
by \citet{1995A&AS..114..393H} was successfully applied to the
study of the pulsating triple system DG Leo. It allowed to
confirm the previously derived parameters of the spectroscopic
orbit by \citet{1977PASP...89..216F} and to increase the accuracy.
Combining astrometric measurements with our new radial-velocity
measurements, we were further able to compute for the first time
the relative orbit of the visual system AB. However, we stress
the fact that speckle and RV measurements at periastron would be
invaluable. We would like therefore to be informed about such
data, if existing, as these measurements should greatly increase
the precision of the present solution and it could further enable
the determination of the stellar masses of all 3 components in an
independent way.

A detailed chemical analysis was performed on the disentangled
spectra of the components {with} the Kurucz LTE model atmospheres
combined with the {\sc synspec} computer codes. All three
components appear to have the same fundamental parameters
(effective temperature and surface gravity) but DG Leo B was found
to have a solar-like chemical composition, while the two
components belonging to the close spectroscopic binary show an
abundance pattern typical of Am stars. A different level of
underabundance for scandium and more strongly for calcium is noted
between the two close components (Aa and Ab). These differences
probably reflect the sensitivity of the diffusion processes to
issues such as rotation, mass loss and evolution. We further
noticed that, while the spectroscopic orbit is circularized, the
apparent rotation of both A components is {not yet} synchronized
with the orbital motion. DG Leo is therefore a very interesting
clue for understanding issues such as rotation synchronization
and orbital circularization.

Our high--time--resolution spectroscopic data confirm the
existence of line--profile variations (LPVs) caused by pulsation
in component B with time scales very close to those recently
detected by photometric studies \citep{phot1}. The analysis and
the mode identification of these LPVs will be combined with the
multicolour photometric observations, with the goal to provide a
better understanding of the non-trivial link existing between
chemical composition, multiplicity and pulsation.

\section*{Acknowledgements}

We thank P. North for providing us with updated atomic data and
spectral line lists. We thank G. Alecian and J.-P. Zahn for an
interesting discussion about diffusion, rotation and
synchronization. We are grateful to P. Mathias, P. North \& E.
Oblak for their help in gathering new radial-velocity data. We
wish to acknowledge the referee, R. Griffin, for his suggestions
which greatly improved the readability of the paper. YF and PL
acknowledge funding from the Belgian Federal Science Policy
(Research project MO/33/007) and from the FNRS (Travel grant to
Dubrovnik). HH acknowledges support from the IAP P5/36 project of
the Belgian Federal Science Policy.

\bibliographystyle{aa}
\bibliography{dgleo5}

\end{document}